\documentclass[a4paper,11pt]{article}
\usepackage{
a4wide,
amsmath,
amsfonts,
amssymb,
ascmac,
bm,
braket,
cite,
comment,
enumerate,
here,
mathtools,
ulem,
}
\usepackage[]{physics}
\usepackage{color,graphicx}
\usepackage[labelsep=quad]{caption}
\usepackage[labelsep=quad,labelformat=empty]{subcaption}
\usepackage{authblk}

\makeatletter
\@addtoreset{equation}{section}
\def\theequation{\thesection.\arabic{equation}}
\makeatother

\newcommand{\figref}[1]{Fig.~\ref{#1}}

\newcommand{\e}[1]{{\mathrm{e}^{#1}}}

\newcommand{\eps}{\epsilon}
\newcommand{\veps}{\varepsilon}
\newcommand{\pp}{{\prime\prime}}
\newcommand{\fcn}[2]{{{#1}\pqty{#2}}}

\newcommand{\pdd}[1]{{\partial_{#1}}}
\newcommand{\lie}[1]{{\mathcal{L}_{#1}}}
\newcommand{\cdd}[1]{{\nabla_{#1}}}
\newcommand{\mj}{{Maxwell--J\"{u}ttner}}

\newcommand{\opop}[2]{\pqty{{#1},{#2}}}

\newcommand{\clop}[2]{\left[{{#1},{#2}}\right)}
\newcommand{\clcl}[2]{\bqty{{#1},{#2}}}

\newcommand{\nt}{\notag\\}

\definecolor{DarkMagenta}{rgb}{0.54,0,0.54}

\definecolor{DarkBlue}{rgb}{0,0,0.7}

\definecolor{DarkRed}{rgb}{0.65,0,0}

\begin{document}
\title{
  Einstein--Vlasov system with equal-angular momenta in $\text{AdS}_5$
}
\author{Hiroki Asami\footnote{Email:\href{mailto:asami.hiroki.b3@s.mail.nagoya-u.ac.jp}{asami.hiroki.b3@s.mail.nagoya-u.ac.jp}}}
\author{Chul-Moon Yoo\footnote{Email:\href{mailto:yoo.chulmoon.k6@f.mail.nagoya-u.ac.jp}{yoo.chulmoon.k6@f.mail.nagoya-u.ac.jp}}}
\author{Ryo Kitaku\footnote{Email:\href{mailto:kitaku.ryo.f4@s.mail.nagoya-u.ac.jp}{kitaku.ryo.f4@s.mail.nagoya-u.ac.jp}}}
\author{Keiya Uemichi\footnote{Email:\href{mailto:uemichi.keiya.j4@s.mail.nagoya-u.ac.jp}{uemichi.keiya.j4@s.mail.nagoya-u.ac.jp}}}
\affil{Division of Particle and Astrophysical Science, Graduate School of Science, Nagoya University, Nagoya 464-8602, Japan}
\setcounter{Maxaffil}{0}
\renewcommand\Affilfont{\itshape\small}
\date{}

\maketitle
\begin{abstract}
  We investigate solutions of the $5$--dimensional rotating Einstein-Vlasov system with an $R \times SU(2) \times U(1)$ isometry group. 
  In a five-dimensional spacetime, there are two independent planes of rotation, 
  thus, considering $U(1)$ symmetry on each rotation plane, we may impose an $R\times U(1) \times U(1)$ isometry to a stationary spacetime. 
  Furthermore, when the values of the two angular momenta are equal to each other, 
  the spatial symmetry gets enhanced to $R\times SU(2)\times U(1)$ symmetry, and the spacetime has a cohomogeneity-1 structure. 
  Imposing the same symmetry to the distribution function of the particles of which the Vlasov system consists, 
  the distribution function can be dependent on three mutually independent and commutative conserved charges for particle motion (energy, total angular momentum on $SU(2)$ and $U(1)$ angular momentum). 
  We consider the distribution function which exponentially depends on the $U(1)$ angular momentum and reduces to the thermal equilibrium state in spherical symmetry.  
  Then, in this paper, we numerically construct solutions of the asymptotically AdS Einstein--Vlasov system.
\end{abstract}

\section{Introduction}
\label{sec:introduction}
General relativistic self-gravitating collisionless many-particle systems, which are often called Einstein-Vlasov systems, have long been investigated in astrophysics.
In particular, spherically symmetric systems have been studied in detail, 
and a great deal of research has been done on the existence of solutions~\cite{cmp/1104251959,Rein:1993ix,Andreasson:2014ina,Andreasson:2015agw} and the stability of the systems~\cite{Andreasson:2006gx,HAD_I__2013,Hadzic:2018bkd,Gunther:2020mvb,Gunther:2021pbf, Had_i__2021} (see also a review \cite{Andreasson:2011ng}).
To realize a specific configuration of the Einstein-Vlasov system, 
it is necessary to impose an appropriate ansatz on the distribution function. 
For example, to give a static configuration surrounding a black hole, we have to assume a distribution with a lower cutoff of the angular momentum~\cite{Andreasson:2021lsh}.
If we assume a {\mj} distribution, we can realize a thermal equilibrium state of the self-gravitating system. 
However, there are no thermal equilibrium states with finite mass in asymptotically flat spacetimes because gravity is a long-range interaction. 
Antonov investigated the non-relativistic many-particle systems with finite mass by introducing an adiabatic wall confining the particles~\cite{antonov1962}, 
and the properties of those systems have been generalized and investigated in detail by Lynden-Bell and Wood~\cite{BellWood1968}.

The thermodynamical instability of self-gravitating systems is often called the gravothermal catastrophe which also applies to relativistic cases. 
It should be noted that, however, since the thermal equilibrium states have infinite mass with a vanishing cosmological constant, 
the analyses of the gravothermal catastrophe in asymptotically flat spacetime always require an artificial wall to confine the system. 
On the other hand, in a system with a negative cosmological constant, the AdS barrier confines the particle system, allowing it to naturally be in a thermal equilibrium state.
Some of the authors have constructed a thermal equilibrium state of such a system confined by the AdS barrier under the assumption of static spherical symmetry and analyzed its stability\cite{Asami_2021,Asami:2022bqc}.

Properties of the Einstein-Vlasov system in static and spherically symmetric cases have been intensively studied, 
but there are still few studies on systems with rotation\cite{1993ApJ...419..622S,1993ApJ...419..636S,Andreasson:2012cq,Ames:2016coj,Thaller:2019skf}. 
  The properties of a system, such as stability, generally depend on the presence of angular momentum because the total angular momentum may prevent the system from collapsing.
  In the case of a self-gravitating many-particle system, it is expected that the instability associated with gravitational collapse may be
    inhibited by the angular momentum of the system.
However, the existence of the non-zero angular momentum inevitably reduces the spacetime symmetry, and the analyses become much more difficult.  
To avoid this technical difficulty, we focus on $5$--dimensional spacetimes because it is known that the spacetime can have a cohomogeneity-1 structure even with non-zero angular momentum in $5$--dimensional spacetimes. 
That is, we can construct a spacetime with finite angular momentum solving a set of ordinary differential equations for unknown variables depending only on a radial coordinate. 
More specifically, in a $5$--dimensional spacetime, if the values of the angular momenta on the two independent rotation planes are equal to each other, 
the spacetime symmetry can be enhanced to $R \times SU(2) \times U(1)$. 
The corresponding black hole solution is called the Myers-Perry (AdS) black hole with equal angular momenta\cite{MYERS1986304,Hawking:1998kw,Gibbons:2004uw,Gibbons:2004js,Kunduri:2006qa}.
Due to its high symmetry, this spacetime is often used in the analyses of gravitational perturbations and phenomena specific to rotating systems.

In this paper, we consider rotating solutions for the Einstein-Vlasov system with an $R \times SU(2) \times U(1)$ isometry group appropriately setting the distribution function of the Vlasov field.  
A negative cosmological constant is introduced for the realization of the stationary solutions with finite total mass and angular momentum. 
In general, a resultant spacetime has an asymptotically locally AdS structure, whose spatial geometry is given by the foliation of squashed $S^3$ hyper-surfaces even at spatial infinity.  
The squashing parameter at spatial infinity can be set to zero by tuning the boundary condition at the center, and the spacetime can be asymptotically AdS without squashing of $S^3$ at spatial infinity.  
We note that a similar situation has been reported in the vacuum cases~\cite{Murata:2009jt}. 

The purpose of introducing a negative cosmological constant is not only for the construction of a physical solution with finite mass. 
Recently, asymptotically AdS spacetimes have attracted much attention in the context of the AdS/CFT correspondence~\cite{Maldacena_1999,Gubser_1998,witten1998anti} and the gravitational turbulent instability~\cite{Bizo__2011}. 
The conditions for the onset of the instability have not yet been clarified~\cite{Dias:2012tq,Maliborski:2013jca,Buchel:2013uba,Balasubramanian:2014cja,Balasubramanian:2015uua,Bizon:2015pfa,Dimitrakopoulos:2015pwa,Green:2015dsa,Garfinkle:2011hm,Jalmuzna:2011qw,Bizon:2017yrh,Craps:2014vaa,Craps:2014jwa,Evnin:2021buq}, 
and there are still few clues to the final state. 
The system provided in this paper may be treated as a macroscopic model of the final state of a system complicated by turbulent phenomena, 
and we expect that our analyses will be helpful to get useful insights into the final state of turbulent instability.     
In addition, the instability in the Einstein-Vlasov system is also discussed in Refs.~\cite{Moschidis:2017llu,Gunther:2020mvb,Moschidis:2017lcr}, and  
the possible relation with the Hawking-Page transition has been pointed out in Refs.~ \cite{Vaganov:2007at,Hammersley:2007ahw}.
Another related phenomenon is the superradiant instability of rotating black holes in asymptotically AdS spacetime, for which the existence of finite angular momentum is essential. 
The final fate of the superradiant instability has not been also clarified yet. 
In order to approach the superradiant instability through the construction of a macroscopic model with the Einstein-Vlasov system, 
the introduction of finite angular momentum is a necessary step to be performed. 

This paper is organized as follows.
In section \ref{sec:setup}, we provide the metric ansatz of the spacetime with an $R \times SU(2) \times U(1)$ isometry group.
We also define the conserved quantities along the geodesic of a particle and list the conditions which we impose on the metric functions for technical and practical reasons.  
In section \ref{sec:physical_quantities}, introducing a specific form of the distribution function, we show the explicit forms for the energy-momentum tensors (detailed calculations are shown in App.~\ref{sec:int_mom}).
We write down the Einstein field equations for our system in section \ref{sec:einstein} and numerically solve them in section \ref{sec:results}.
Section \ref{sec:conclusion} is devoted to a summary and conclusion.

Throughout this paper, we use the geometrized units in which both the speed of light and gravitational constant in $5$--dimension are unity, $c=G=1$.

\section{Metric Ansatz and conserved quantities}
\label{sec:setup}
\subsection{Metric ansatz with an $R \times SU(2) \times U(1)$ isometry group}
We start with the following form of the metric:
\begin{align}
  \bm{g} \coloneqq -\e{2\fcn{\mu}{r}}\dd{t}^2 +\e{2\fcn{\nu}{r}}\dd{r}^2 +\frac{r^2}{4}\bqty{\pqty{\bm{\sigma}^1}^2 +\pqty{\bm{\sigma}^2}^2 +\pqty{\bm{\sigma}^3}^2} +\fcn{h}{r}\pqty{\dd{t}-\frac{\fcn{a}{r}}{2}\bm{\sigma}^3}^2,
  \label{eq:metric}
\end{align}
where $\bm{\sigma}^i$ are one-forms defined as
\begin{subequations}
  \begin{align}
    \bm{\sigma}^1 &\coloneqq -\sin{\phi}\dd{\theta} +\sin{\theta}\cos{\phi}\dd{\psi},\\
    \bm{\sigma}^2 &\coloneqq -\cos{\phi}\dd{\theta} -\sin{\theta}\sin{\phi}\dd{\psi},\\
    \bm{\sigma}^3 &\coloneqq \dd{\phi} +\cos{\theta}\dd{\psi},
  \end{align}
  \label{eq:sigma_form}%
\end{subequations}
satisfying the Maurer-Cartan equation $\dd{\bm{\sigma}^i} +\tfrac{1}{2}{\eps^i}_{jk}\bm{\sigma}^j\wedge\bm{\sigma}^k=0$.
The ranges of the coordinate variables are given by 
$t\in\opop{-\infty}{\infty}$, $r\in\clop{0}{\infty}$, $\theta\in\clcl{0}{\pi}$, $\phi\in\clop{0}{4\pi}$ and $\psi\in\clop{0}{2\pi}$.
The third term in the metric 
\begin{align}
  \bm{\gamma} 
  \coloneqq &\frac{1}{4}\bqty{\pqty{\bm{\sigma}^1}^2 +\pqty{\bm{\sigma}^2}^2 +\pqty{\bm{\sigma}^3}^2} \nt
  = &\frac{1}{4}\pqty{\dd{\theta}^2 +\dd{\phi}^2 +\dd{\psi}^2 +2\cos\theta\dd{\phi}\dd{\psi}}.
  \label{eq:metric_sphere}
\end{align}
describes the metric on $S^3$. 

The three-sphere $S^3$ has two sets of $SU(2)$ generators $\Bqty{\bm{\xi}_i}_{i\in\Bqty{1,2,3}}$ and $\Bqty{\bm{\sigma}_i}_{i\in\Bqty{1,2,3}}$ written in the form:
\begin{subequations}
  \begin{align}
    \bm{\xi}_1 &= -\sin{\psi}\pdd{\theta} +\csc{\theta}\cos{\psi}\pdd{\phi} -\cot{\theta}\cos{\psi}\pdd{\psi},\\
    \bm{\xi}_2 &= -\cos{\psi}\pdd{\theta} -\csc{\theta}\sin{\psi}\pdd{\phi} +\cot{\theta}\sin{\psi}\pdd{\psi},\\
    \bm{\xi}_3 &= \pdd{\psi},
  \end{align}
  \label{eq:killing_xi}%
\end{subequations}
and 
\begin{subequations}
  \begin{align}
    \bm{\sigma}_1 &= -\sin{\phi}\pdd{\theta} -\cot{\theta}\cos{\phi}\pdd{\phi} +\csc{\theta}\cos{\phi}\pdd{\psi},\\
    \bm{\sigma}_2 &= -\cos{\phi}\pdd{\theta} +\cot{\theta}\sin{\phi}\pdd{\phi} -\csc{\theta}\sin{\phi}\pdd{\psi},\\
    \bm{\sigma}_3 &= \pdd{\phi}.
  \end{align}
  \label{eq:killing_sigma}%
\end{subequations}
The $SU(2)$ generators $\Bqty{\bm{\xi}_i,\bm{\sigma}_i}$ satisfy the following commutation relations: 
\begin{align}
  \bqty{\bm{\xi}_i,\bm{\xi}_j} = {\eps_{ij}}^k\bm{\xi}_k \qc
  \bqty{\bm{\sigma}_i,\bm{\sigma}_j} = {\eps_{ij}}^k\bm{\sigma}_k \qc
  \bqty{\bm{\xi}_i,\bm{\sigma}_j} = \bqty{\bm{\sigma}_i,\bm{\xi}_j} = 0,
  \label{eq:commutator_on_sphere}
\end{align}
or equivalently 
\begin{align}
  \lie{\bm{\xi}_i}\bm{\xi}^j = {{\eps_i}^j}_k \bm{\xi}^k \qc
  \lie{\bm{\sigma}_i}\bm{\sigma}^j = {{\eps_i}^j}_k \bm{\sigma}^k \qc
  \lie{\bm{\xi}_i}\bm{\sigma}^j = 0.
  \label{eq:lie_forms}
\end{align}
The Killing vectors $\Bqty{\bm{\xi}_i}$ and $\bm{\sigma}_3$ on the three-sphere are also the Killing vectors on the spacetime due to Eq.~\eqref{eq:lie_forms}.
On the other hand, neither $\bm{\sigma}_1$ nor $\bm{\sigma}_2$ is the Killing vector on the spacetime due to the last term in Eq.~\eqref{eq:metric} unless $\fcn{a}{r} = 0$ everywhere.
Since the vector $\bm{\sigma}_3$ generates a $U(1)$ isometry group and the metric has a time-like Killing vector $\bm{\eta} = \pdd{t}$, 
the spacetime has the $R_t \times SU(2)_{\bm{\xi}} \times U(1)_{\bm{\sigma}}$ isometry group. 
The black hole solutions which have the same symmetry are known as the Myers--Perry black holes with equal angular momenta~\cite{Gibbons:2004uw,Gibbons:2004js}. 
It would be worth noting that the angular coordinates $\theta$, $\phi$ and $\psi$ are related to the Hopf coordinates $\tilde \theta$, $\tilde \phi$ and $\tilde \psi$ 
through $\theta=2\tilde \theta$, $\psi=-\tilde \phi+\tilde \psi$ and $\phi=\tilde \phi+\tilde \psi$. 
Then the two equal angular momenta are the conserved charges associated with the Killing vectors $\partial_{\tilde \phi}$ and $\partial_{\tilde \psi}$.

In this paper, to avoid possible technical problems, we focus on the cases satisfying the following four conditions: 
\begin{enumerate}
  \item Non-degeneracy: $\det\bm{g}<0$, \label{cond:non-degeneracy} 
  \item No-horizon: $\bm{r}\cdot\bm{r} > 0$, \label{cond:no-horizon}
  \item Time-like Killing vector exists everywhere: $\bm{\eta}\cdot\bm{\eta} < 0$, \label{cond:no-ergo}
  \item Time-like unit normal exists everywhere: $\bm{n}\cdot\bm{n} < 0$, \label{cond:foliation}
\end{enumerate}
where $\bm{r} \coloneqq \abs{g^{rr}}^{-\frac{1}{2}}\dd{r}$ and $\bm{n} \coloneqq -\abs{g^{tt}}^{-\frac{1}{2}}\dd{t}$ are the unit normal to the $r={\rm const.}$ and the $t={\rm const.}$ hyper-surfaces, respectively.
The condition \ref{cond:no-ergo} implies that the spacetime has no ergo-region\footnote{In Ref.~\cite{Ames:2016coj}, the solution with an ergo-region is constructed in a 4--dimensional asymptotically flat spacetime. }.
The conditions \ref{cond:non-degeneracy} - \ref{cond:foliation} yield
\begin{align}
  \e{2\mu(r)}-h(r),\ \e{2\nu(r)},\ F(r),\ G(r) > 0,
  \label{eq:metricConditions}
\end{align}
for any $r > 0$, where $\fcn{G}{r} \coloneqq r^2+a^2h$ and $\fcn{F}{r} \coloneqq \e{2\mu}G-hr^2$.

\subsection{Conserved quantities for the geodesic motion}
The symmetry of the spacetime indicates the existence of conserved quantities for the geodesic motion in the form of the inner product between the Killing vector and the momentum of a particle.  
Then we can consider the following mutually independent conserved quantities for the metric ~\eqref{eq:metric}: 
\begin{subequations}
  \begin{align}
    \veps 
    &= -\bm{p}\cdot\bm{\eta} = -p_t, \\
    {J_{\xi_1}} 
    &= \bm{p}\cdot\bm{\xi}_1 = -p_{\theta}\sin{\psi} +p_{\phi}\csc{\theta}\cos{\psi} -p_{\psi}\cot{\theta}\cos{\psi}, \\
    {J_{\xi_2}} 
    &= \bm{p}\cdot\bm{\xi}_2 = -p_{\theta}\cos{\psi} -p_{\phi}\csc{\theta}\sin{\psi} +p_{\psi}\cot{\theta}\sin{\psi}, \\
    {J_{\xi_3}} 
    &= \bm{p}\cdot\bm{\xi}_3 = p_{\psi}, \\
    {j_{\sigma}} 
    &= \bm{p}\cdot\bm{\sigma}_3 = p_\phi,
  \end{align}
  \label{eq:def_conserved_quantities}%
\end{subequations}
where $\bm{p}$ is the momentum of the particle and `$\cdot$' denotes the inner product concerning $\bm{g}$. 
The total angular momentum $J_\xi \ge 0$ of the $SU(2)_{\bm{\xi}}$ sector can be defined as 
\begin{align}
  {J_\xi}^2 \coloneqq \sum_{i=1}^3 {J_{\xi_i}}^2 = \frac{1}{4}\gamma^{\mu\nu}p_\mu p_\nu,
  \label{eq:def_J_xi}
\end{align}
where we have defined $\gamma^{\mu\nu}$ as $\gamma^{\mu\nu} \coloneqq 4\sum_i {\sigma_i}^\mu {\sigma_i}^\nu$. 
Then $\veps$, $j_\sigma$ and $J_\xi$ are mutually commutative conserved charges and can be used as independent coordinates in the phase space. 

The momentum of the particle must satisfy the on-shell condition: $p^2+m^2=0$, which can be rewritten as follows:
\begin{align}
  \frac{\e{-2\nu}G}{F}\pqty{\veps-\frac{2ah}{G}j_\sigma}^2 -\e{-2\nu}\bqty{m^2+\frac{4}{r^2}\pqty{{J_\xi}^2 -\frac{a^2h}{G}{j_\sigma}^2}} = \pqty{p^r}^2. 
  \label{eq:on-shell_condition}
\end{align}
We note that the left-hand side of Eq.~\eqref{eq:on-shell_condition} can be regarded as the effective potential for the geodesic motion. 
Here we impose that the momentum is future pointing $\bm n\cdot \bm p < 0$. 
Then the positivity of the local energy of the particle is ensured:
\begin{align}
  \veps-\frac{2ah}{G}j_\sigma > 0.
  \label{eq:positivity_condition}
\end{align}
The allowed region in the momentum space for the particle is the subspace satisfying the conditions~\eqref{eq:on-shell_condition} and \eqref{eq:positivity_condition}.

\section{Model and physical quantities}
\label{sec:physical_quantities}

\subsection{Distribution function and the energy-momentum tensor}

Considering a collisionless many-particle system, the particles follow the geodesic motion.
Therefore the distribution function $f$ satisfies the Vlasov (collisionless Boltzmann) equation:
\begin{align}
  p^\mu\cdd{\mu}f = p^\mu\pdv{f}{x^\mu} -{\Gamma^i}_{\mu\nu}p^\mu p^\nu\pdv{f}{p^i}= 0,
  \label{eq:vlasov_eq}
\end{align}
which implies the conservation of the distribution function along the geodesic.
If the distribution function is written in the form
\begin{align}
  f = f(\veps, J_\xi, j_\sigma),
  \label{eq:dist_ansatz}
\end{align}
the Vlasov equation~\eqref{eq:vlasov_eq} is automatically satisfied because $\veps$, $J_\xi$ and $j_\sigma$ are conserved quantities along the geodesic. 

In this paper, as a simple specific model, we assume the following distribution function
\begin{align}
  \fcn{f}{\veps,j_\sigma} = \exp\bqty{\alpha-\beta(\veps-\Omega j_\sigma)},
  \label{eq:mjlike_dist}
\end{align}
with constants $\alpha \in \mathbb{R}$, $\beta>0$ and $\Omega>0$.
The distribution function~\eqref{eq:mjlike_dist} reduces to the {\mj} distribution, which describes the relativistic thermal equilibrium states in static cases, with $\Omega=0$.
Thus the system with this distribution function can be regarded as an extension of the thermal equilibrium state to the system with a finite angular momentum.
The {\mj} distribution is derived by extremizing the entropy of the system fixing the total mass and the total particle number in static cases.
On the other hand, in rotating cases, it is not clear how to determine the most probable rotating state.
Therefore the functional form \eqref{eq:mjlike_dist} should be regarded as a working assumption for $\Omega\neq 0$.

Given a one-particle distribution function $f$, the energy-momentum tensor $T_{\mu\nu}$ is given by
\begin{align}
  T_{\mu\nu} \coloneqq \int \dd{V_p}p_\mu p_\nu f(x^\mu,p^i),
  \label{eq:def_emt}
\end{align}
where $\dd{V_p}$ is the integral measure in the momentum space:
\begin{align}
  \dd{V_p} \coloneqq -\frac{2\fcn{\delta}{p^2+m^2}\fcn{\Theta}{\veps-2ah j_\sigma/G}}{\sqrt{-\det \bm{g}}}\dd{p_t}\wedge\dd{p_r}\wedge\dd{p_\theta}\wedge\dd{p_\phi}\wedge\dd{p_\psi}.
  \label{eq:def_momentum_measure}
\end{align}
Here the delta function $\delta$ and the Heaviside's step function $\Theta$ describe the conditions~\eqref{eq:on-shell_condition} and \eqref{eq:positivity_condition}, respectively.

To obtain numerical solutions, we have to know the local expressions of energy-momentum tensor by integrating over the momentum space.
In spherical cases, we can perform the integration analytically and obtain explicit expressions because the metric is diagonal and the on-shell condition is simple as shown in Refs.~\cite{Asami_2021,Asami:2022bqc}.
On the other hand, in rotating cases, we cannot perform analytical integrations due to the off-diagonal components of the metric and the cross term in Eq.~\eqref{eq:on-shell_condition}.
As is shown in App.~\ref{sec:int_mom}, however, we can reduce it to one-dimensional integrals of the normalized energy under the ansatz~\eqref{eq:mjlike_dist}. 
That is, we can obtain the values of the energy-momentum tensor at each radial coordinate $r$ by performing one-dimensional numerical integration. 
The explicit expressions are shown in App.~\ref{sec:int_mom}.

\subsection{Komar integral and total particle number}
Before solving the Einstein equations, let us define global conserved quantities characterizing the system in terms of the Komar integrals.

We define the total mass of the system by the Komar mass:
\begin{align}
  M_K \coloneqq \int\dd{\Sigma} n_\mu {T^\mu}_{\nu} \eta^\nu,
  \label{eq:def_komar_mass}
\end{align}
where $\dd{\Sigma} = 2\pi^2 r^2 \e{\nu}\sqrt{G}\dd{r}$ is the invariant volume element on a $t={\rm const.}$ hyper-surface. 
In our formulation, Eq.~\eqref{eq:def_komar_mass} becomes
\begin{align}
  M_K = 2\pi^2\int\dd{r} \frac{r^2\e{\nu}}{\sqrt{F}}(GT_{tt} + 2ahT_{t\phi}).
\end{align}
We note that the Komar mass for the distribution~\eqref{eq:def_komar_mass} can be finite due to the potential wall for massive particles associated with a negative cosmological constant $\Lambda<0$. 

We also define the total angular momentum of the system by
\begin{align}
  J_\phi \coloneqq -\int\dd{\Sigma} n_\mu {T^\mu}_{\nu} {\sigma_3}^\nu.
  \label{eq:def_komar_am}
\end{align}
Similarly to the Komar mass, in our formulation, we obtain the expression:
\begin{align}
  J_\phi = -2\pi^2\int\dd{r} \frac{r^2\e{\nu}}{\sqrt{F}}(GT_{t\phi} + 2ahT_{\phi\phi}).
\end{align}

Since the Komar integrals associated with the Killing vectors conserve, we can choose them as quantities characterizing the system.

\section{Einstein equations}
\label{sec:einstein}
\subsection{Field equations}
For a 5--dimensional spacetime, the Einstein field equations $G_{\mu\nu} +\Lambda g_{\mu\nu} = 8\pi T_{\mu\nu}$ with $\Lambda<0$ can be rewritten as
\begin{align}
  R_{\mu\nu} = 8\pi \pqty{{T}_{\mu\nu} -\frac{{T}}{3} g_{\mu\nu}-\frac{g_{\mu\nu}}{2\pi L^2}}, 
  \label{eq:einstein}
\end{align}
where $L\coloneqq \sqrt{-6/\Lambda}$ is the AdS radius.

For numerical analyses, we rewrite the equations by dimensionless variables.
We define $\ell \coloneqq (8\pi T_{tt}(0))^{-1/2}$ and consider normalized variables $x\coloneqq r/\ell$ and $\lambda \coloneqq L/\ell$.
As alternative functions, we introduce the following dimensionless metric functions:
\begin{align}
  y_1(x) \coloneqq \frac{ah}{\ell} \qc y_2(x) \coloneqq \frac{a^2h}{\ell^2} \qc y_3(x) \coloneqq \e{2\mu}-h \qc y_4(x) \coloneqq \e{2\nu}.
  \label{eq:alternative_metric_func}
\end{align}
We also define $f_1(x)\coloneqq G/\ell^2 = x^2 +y_2$ and $f_2(x) \coloneqq F/\ell^2 = {y_1}^2 +f_1 y_3$ for a notational simplicity.

Defining
\begin{align}
  s_{\mu\nu} \coloneqq \frac{1}{T_{tt}(0)}\bqty{{T}_{\mu\nu} -\frac{T}{3}g_{\mu\nu} -\frac{1}{2\pi L^2}g_{\mu\nu}},
\end{align}
relevant components of the Einstein equations are as follows:
\begin{subequations}
  \begin{align}
    {y_1}^\pp
    &= \frac{2x({{{y_1}^\prime}}{y_3} -{{{y_3}^\prime}}{y_1})}{f_2} +\frac{{f_1}{{{y_1}^\prime}}{{{y_3}^\prime}} +{{{y_1}^\prime}}^2{y_1} +{{{y_1}^\prime}}{{{y_2}^\prime}}{y_3} -{{{y_2}^\prime}}{{{y_3}^\prime}}{y_1}}{f_2} \notag\\
    &\hspace{20pt} +4s_{t\phi}{y_4} +\frac{{{{y_1}^\prime}}({f_3}{y_4}-2)}{2x} +\frac{2{{{y_1}^\prime}}{y_2}{y_4}}{x^3} +\frac{4{f_1}{y_1}{y_4}}{x^4}, \\
    {y_2}^\pp
    &= \frac{4x({{{y_1}^\prime}}{y_1} +{{{y_2}^\prime}}{y_3})}{f_2} +\frac{{{{y_2}^\prime}}^2{y_3} -{f_1}{{{y_1}^\prime}}^2 +2{{{y_1}^\prime}}{{{y_2}^\prime}}{y_1}-4({y_1}^2 +{y_2}{y_3})}{f_2} \notag\\
    &\hspace{20pt} +{f_4}{y_4} +\frac{{{{y_2}^\prime}}({f_3}{y_4}-2)}{2x} +\frac{2{y_2}{y_4}}{x^2}\pqty{6+\frac{{{y_2}^\prime}}{x} +\frac{2{y_2}}{x^2}}, \\
    {y_3}^\pp
    &=  \frac{{f_1}{{{y_3}^\prime}}^2 -{{{y_1}^\prime}}^2{y_3} +2{{{y_1}^\prime}}{{{y_3}^\prime}}{y_1}}{f_2} \notag\\
    &\hspace{20pt} +2s_{tt}{y_4} +\frac{{{{y_3}^\prime}}({f_3}{y_4}-2)}{2x} +\frac{2{{{y_3}^\prime}}{y_2}{y_4}}{x^3} -\frac{4{y_1}^2{y_4}}{x^4},\\
    {y_4}^\prime
    &= {y_4}\bqty{\frac{2x{y_3}}{f_2} +\frac{{f_1}{{{y_3}^\prime}} +2{{{y_1}^\prime}}{y_1} +{{{y_2}^\prime}}{y_3}}{f_2} +\frac{{f_3}{y_4}+2}{x} +\frac{4{y_2}{y_4}}{x^3}},
  \end{align}
  \label{eq:einstein_num}%
\end{subequations}
where $\ ^\prime\coloneqq \dv*{}{x}$, $f_3 \coloneqq -4(1+s_{\phi\phi})+\gamma^{\mu\nu}s_{\mu\nu}$ and $f_4 \coloneqq -12s_{\phi\phi} + \gamma^{\mu\nu}s_{\mu\nu}$. 
We can show the other components of the Einstein equations are redundant by using Eqs.~\eqref{eq:einstein_num} and continuity equations.

\subsection{Asymptotic behaviors and boundary conditions}
Let us consider the boundary conditions for Eqs.~\eqref{eq:einstein_num} to realize asymptotically AdS spacetimes.
We can rewrite the metric as
\begin{align}
  \ell^{-2}\bm{g} = -\frac{f_2}{f_1}\dd{t}^2 +{y_4}\dd{x}^2 +\frac{x^2}{4}\bqty{\pqty{\bm{\sigma}^1}^2 +\pqty{\bm{\sigma}^2}^2 +\frac{f_1 }{x^2}\pqty{\bm{\sigma}^3-\frac{2y_1}{f_1}\dd{t}}^2}.
  \label{eq:metric_2}%
\end{align}
Setting $T_{\mu\nu}=0$ and expanding the metric functions in the vicinity of the boundary, we obtain the following asymptotic solutions:
\begin{subequations}
  \begin{align}
    \frac{f_2}{f_1} 
    &= \frac{x^2}{\lambda^2} + 1 -s +\order{x^{-1}}, \\
    \frac{1}{y_4} 
    &= \frac{x^2}{\lambda^2} + 1 -\frac{3}{5}s +\order{x^{-1}}, \\
    \frac{f_1}{x^2} 
    &= 1 +s -2s \pqty{1+s} \frac{\lambda^2}{x^2} +\order{x^{-3}}, \\
    \frac{2y_1}{f_1} 
    &= \omega +\order{x^{-3}},%
  \end{align}%
  \label{eq:expansionInfinity}%
\end{subequations}%
where $s$ and $\omega$ can be regarded as free parameters for the asymptotic solution, and the coordinates $t$ and $r$ are fixed so that the pure AdS metric can be realized for $s = \omega = 0$.  
The leading term $\omega$ of $2y_1/f_1$ can be eliminated by employing the co-rotating coordinate $\hat \phi \coloneqq \phi-\omega t$. 
Then we can introduce one-forms $\hat{\sigma}^i$ at the boundary by replacing $\phi$ with $\hat{\phi}$.
Using these one-forms, we can write
\begin{align}
  \ell^{-2}\bm{g} = -\frac{f_2}{f_1}\dd{t}^2 +{y_4}\dd{x}^2 +\frac{x^2}{4}\bqty{\pqty{\hat{\bm{\sigma}}^1}^2 +\pqty{\hat{\bm{\sigma}}^2}^2 +\frac{f_1 }{x^2}\pqty{\hat{\bm{\sigma}}^3}^2}.
  \label{eq:metric_hat}%
\end{align}
As we can see from Eq.~\eqref{eq:expansionInfinity} and Eq.~\eqref{eq:metric_hat}, the parameter $s$ should vanish so that the spacetime has the asymptotically AdS structure.
We note that the contribution of the finite value of $s$ corresponds to the squashing of $S^3$ at infinity as is reported in Ref.~\cite{Murata:2009jt} for vacuum cases. 

Next, let us investigate the asymptotic behavior around the center by expanding the metric functions as
\begin{subequations}
  \begin{align}
    y_1(x) &\simeq x^2 \bqty{{z_1}^{(0)} + {z_1}^{(1)}x + \frac{{z_1}^{(2)}}{2}x^2 +\order{x^3}}, \\
    y_2(x) &\simeq x^2 \bqty{{z_2}^{(0)} + {z_2}^{(1)}x + \frac{{z_2}^{(2)}}{2}x^2 +\order{x^3}}, \\
    y_3(x) &\simeq {y_3}^{(0)} + {y_3}^{(1)}x + \frac{{y_3}^{(2)}}{2}x^2 +\order{x^3}, \\
    y_4(x) &\simeq {y_4}^{(0)} + {y_4}^{(1)}x + \frac{{y_4}^{(2)}}{2}x^2 +\order{x^3},
  \end{align}
  \label{eq:expand}%
\end{subequations}
with coefficients ${z_1}^{(i)}$, ${z_2}^{(i)}$, ${y_3}^{(i)}$ and ${y_4}^{(i)}$.
As regularity conditions, first, we assume the metric components for the Cartesian coordinates $(t,x_1,x_2,x_3,x_4)$\footnote{
  The spatial Cartesian coordinates can be defined by 
  \begin{align}
    \begin{aligned}
      x_1 &= r\sin\left(\frac{\theta}{2}\right)\cos\left[\frac{1}{2}\left(\phi-\psi\right)\right] \qc 
      x_2 = r\sin\left(\frac{\theta}{2}\right)\sin\left[\frac{1}{2}\left(\phi-\psi\right)\right], \\
      x_3 &= r\cos\left(\frac{\theta}{2}\right)\cos\left[\frac{1}{2}\left(\phi+\psi\right)\right] \qc 
      x_4 = r\cos\left(\frac{\theta}{2}\right)\sin\left[\frac{1}{2}\left(\phi+\psi\right)\right]. 
    \end{aligned}
  \end{align}
}
to be $g_{tt}=-y_3^{(0)}$, $g_{x_ix_i}=1 +\order{x^2}$, and $\order{x^2}$ for other components. 
Then, we require all the Ricci tensor components $R_{\mu\nu}$ to be $R_{\mu\nu}={\rm const.} +\order{x}$. 
As a result, we obtain 
\begin{align}
  {z_1}^{(0)} = {z_1}^{(1)} = {z_2}^{(0)} = {z_2}^{(1)} = {y_3}^{(1)} = 0 \qc {y_4}^{(0)} = 1,
  \label{eq:bc_coefficients}
\end{align}
and ${y_3}^{(0)}$ is kept as a free parameter.
As is shown in App.~\ref{sec:bc}, the system has another free parameter ${z_2}^{(2)}$.

When we solve the Einstein equations from the center, in general, we obtain the solution with finite values of $s$ and $\omega$. 
As will be shown later, we obtain a finite value of $s$ at infinity without fine-tuning for the boundary condition at the center. 
In other words, we can set $s=0$ at infinity by tuning ${z_2}^{(2)}$ appropriately as same as Ref.~\cite{Murata:2009jt}.
Therefore, we can construct an asymptotically AdS Einstein-Vlasov system by tuning the parameters $y_3^{(0)}$ and ${z_2}^{(2)}$. 
As a whole, the system has one free parameter $y_3^{(0)}$ corresponding to the linear scaling of the time coordinate, one tuning parameter ${z_2}^{(2)}$ and three physical parameters $(\beta, \Omega, \lambda)$. 

To obtain a solution, we solve Einstein equations~\eqref{eq:einstein_num} with boundary conditions~\eqref{eq:bc_coefficients} at the center and a parameter set $(\beta, \Omega, \lambda)$.
However, for a finite value of $\Omega$, the integral~\eqref{eq:funcIijnu} cannot be performed analytically.
Therefore we need to numerically integrate the integral~\eqref{eq:funcIijnu} at each step of solving the differential equations~\eqref{eq:einstein_num}.
In our numerical simulations, we used 4th-order Runge-Kutta methods in numerical integrations of Eq.~\eqref{eq:funcIijnu} and Eq.~\eqref{eq:einstein_num}.

\section{Results}
\label{sec:results}

\subsection{Static and spherically symmetric cases}
\begin{figure}
  \begin{subfigure}[]{0.5\linewidth}
    \centering
    \includegraphics[width = 72mm]{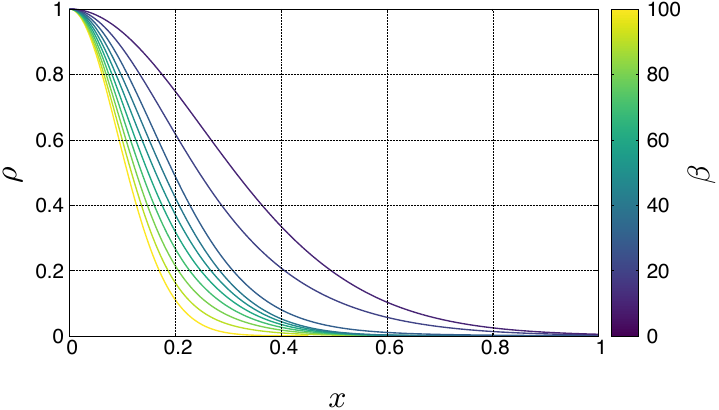}
    \subcaption{$\beta$ dependence for $\lambda=1$}
    \label{fig:Sph_vary_beta}
  \end{subfigure}
  \begin{subfigure}[]{0.5\linewidth}
    \centering
    \includegraphics[width = 72mm]{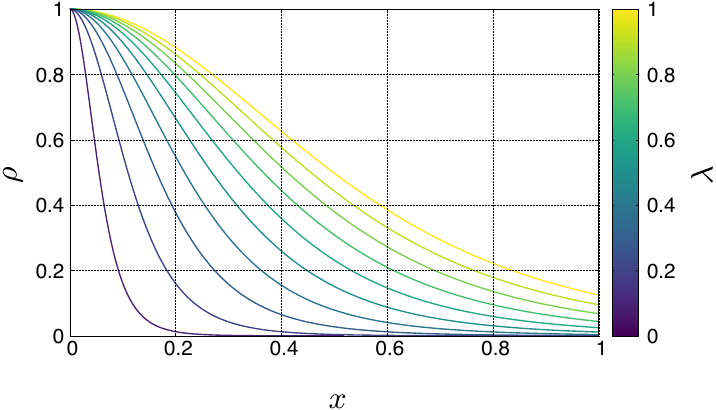}
    \subcaption{$\lambda$ dependence for $\beta=1$}
    \label{fig:Sph_vary}
  \end{subfigure}
  \caption{
    The parameter dependence of the distribution of the energy density for the spherically symmetric cases.
    As the parameter $\beta$ increases, the gravitational potential gets deeper and 
    the system gets more compact.
    As the parameter $\lambda$ decreases, the system gets more compact 
    because the value of $\lambda$ characterizes the radius of the potential wall associated with the negative cosmological constant.
    }
  \label{fig:Sph_density} 
\end{figure}

\begin{figure}
  \begin{subfigure}[]{0.5\linewidth}
    \centering
    \includegraphics[width = 72mm]{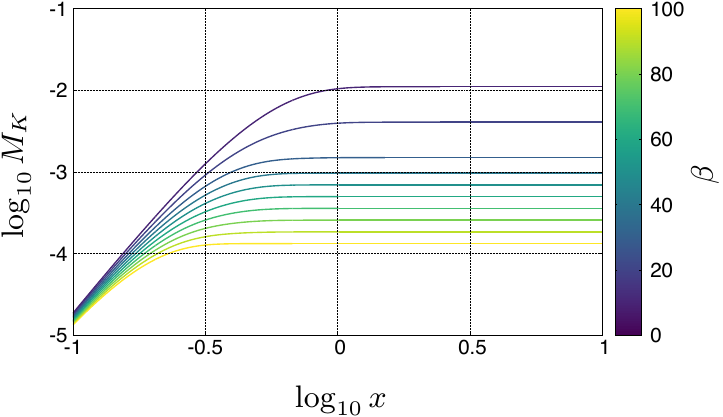}
    \subcaption{$\beta$ dependence for $\lambda=1$}
    \label{fig:Sph_vary_beta_mass}
  \end{subfigure}
  \begin{subfigure}[]{0.5\linewidth}
    \centering
    \includegraphics[width = 72mm]{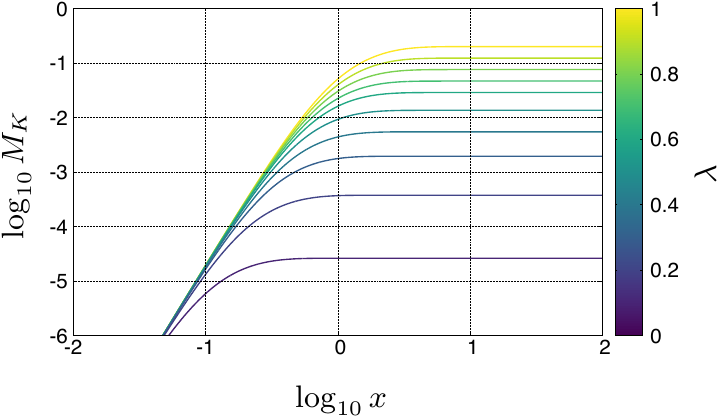}
    \subcaption{$\lambda$ dependence for $\beta=1$}
    \label{fig:Sph_vary_mass}
  \end{subfigure}
  \caption{
    The parameter dependence of the Komar mass for the spherically symmetric cases. 
    The system with relatively small $\beta$ (i.e., high-temperature system) has a large total mass, which 
    implies a deep gravitational potential wall. 
    Since the parameter $\lambda$ works like the size of the system, the Komar mass takes a smaller value for smaller $\lambda$.}
  \label{fig:Sph_mass} 
\end{figure}

First, we investigate the parameter dependence for the spherically symmetric cases with ${z_2}^{(2)}=0$, which corresponds to $s=0$. 
Fig.~\ref{fig:Sph_density} shows the normalized energy density profile for the matter sector $\rho(x)\coloneqq -{T^t}_t/{T}_{tt}(0)$ as a function of $x$ for $\Omega = 0$.
In the spherically symmetric cases, the system is characterized by the two parameters $(\beta,\lambda)$.
The parameter $\beta$ corresponds to the inverse temperature of the system and it characterizes the depth of the gravitational potential. 
Thus, increasing the value of $\beta$ for a fixed value of $\lambda$, 
the density distribution gets sharper and the value of the total mass increases 
as is shown in \figref{fig:Sph_vary_beta} and  \figref{fig:Sph_mass}, respectively. 
\figref{fig:Sph_mass} shows the quasi-local mass given by the Komar integral as a function of $x$.
For all the cases, the mass increases as $M_K \propto x^4$ for $x\ll1$ and asymptotes to a constant value for $x\gg\lambda$.
Since the AdS radius $\lambda$ gives the characteristic radius of the potential wall which confines the Vlasov matter, 
the Komar mass gets larger for large $\lambda$ if the parameter $\beta$ is fixed.

\subsection{Stationary and rotating case}
\begin{figure}
  \begin{subfigure}[]{0.5\linewidth}
    \centering
    \includegraphics[width = 72mm]{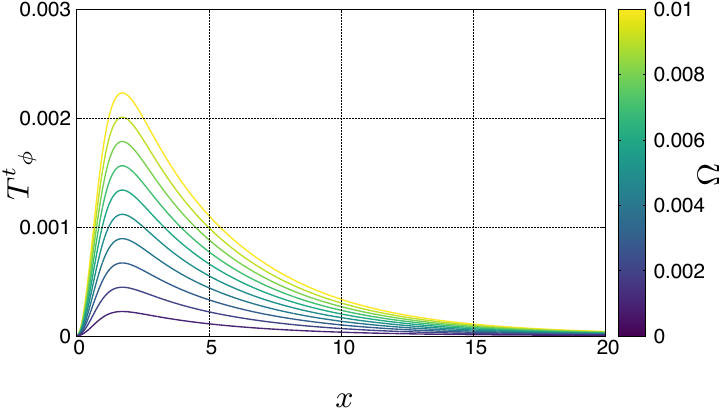}
    \subcaption{Angular momentum density.}
    \label{fig:am_density}
  \end{subfigure}
  \begin{subfigure}[]{0.5\linewidth}
    \centering
    \includegraphics[width = 72mm]{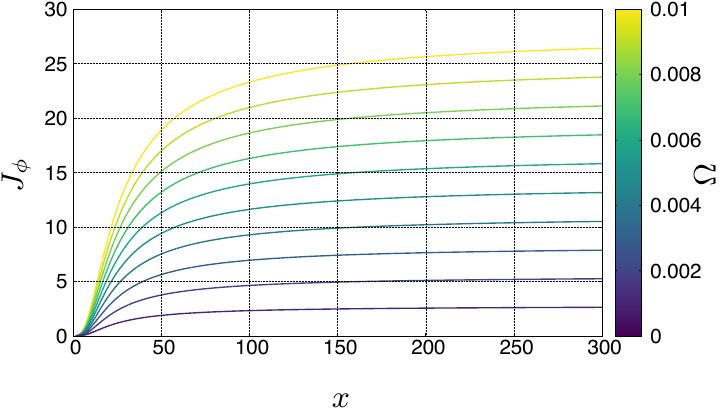}
    \subcaption{Komar angular momentum.}
    \label{fig:komar_am}
  \end{subfigure}
  \caption{
    The angular momentum density $j_\phi = {T^t}_\phi$ and the total angular momentum in the Komar expression for $\beta=0.02$ and $\lambda=10$.
    }
  \label{fig:angular_momentum} 
\end{figure}

We investigate the non-spherically symmetric cases by setting finite values of $\Omega$.
Since the dependence on $\beta$ and $\lambda$ is similar to the spherically symmetric cases,
we focus on the $\Omega$ dependence and the distribution of the angular momentum density of the system.
\figref{fig:angular_momentum} shows the profile of the angular momentum density $j_\phi\coloneqq{T^t}_\phi$ and 
the Komar angular momentum $J_\phi$ for ${z_2}^{(2)} = 0$, $\beta=0.01$ and $\lambda=10$.
For $\Omega\neq0$, the spherical symmetry is broken and the system has a finite angular momentum. 
Since the angular momentum is carried by the Vlasov matter, the distribution of the angular momentum density also starts to decay around the AdS radius $x=\lambda$. 
Therefore, as is shown in \figref{fig:angular_momentum}, the angular momentum density increases as $x$ increases up to a certain value and then it decreases and vanishes asymptotically.

\begin{figure}
  \begin{subfigure}[]{0.5\linewidth}
    \centering
    \includegraphics[width = 72mm]{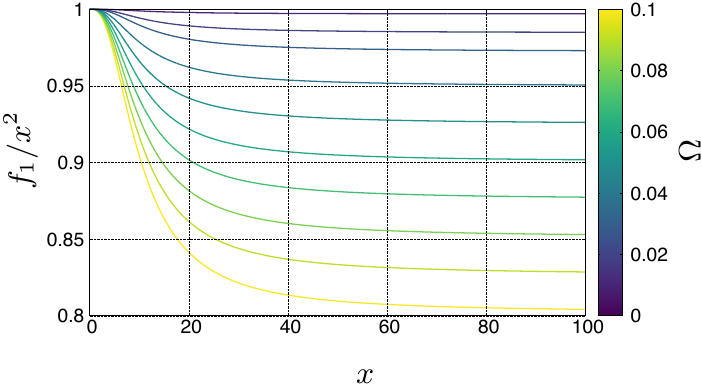}
    \subcaption{The function $f_1/x^2$}
    \label{fig:squashFunction}
  \end{subfigure}
  \begin{subfigure}[]{0.5\linewidth}
    \centering
    \includegraphics[width = 72mm]{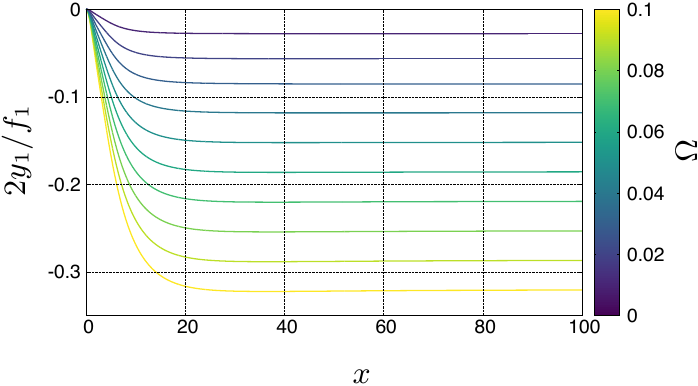}
    \subcaption{The function $2y_1/f_1$}
    \label{fig:draggingFunction}
  \end{subfigure}
  \caption{
    The functions $f_1/x^2$ and $2y_1/f1$ for ${z_2}^{(2)}=0$, $\beta=0.001$ and $\lambda=10$.
    Both of the functions asymptote to constants. 
    }
  \label{fig:functions} 
\end{figure}

\begin{figure}
  \begin{subfigure}[]{0.5\linewidth}
    \centering
    \includegraphics[width = 72mm]{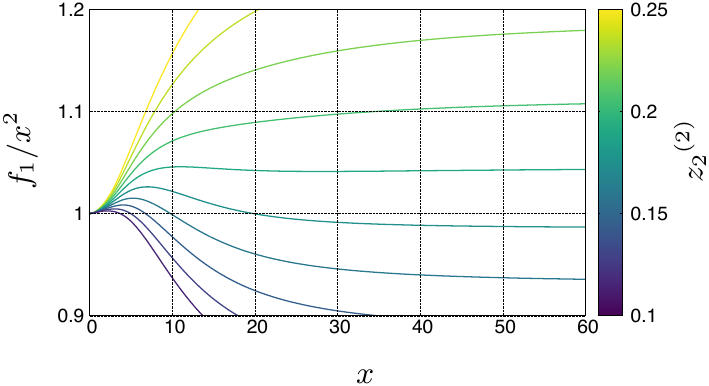}
    \subcaption{The function $f_1/x^2$}
    \label{fig:tuningSquash}
  \end{subfigure}
  \begin{subfigure}[]{0.5\linewidth}
    \centering
    \includegraphics[width = 72mm]{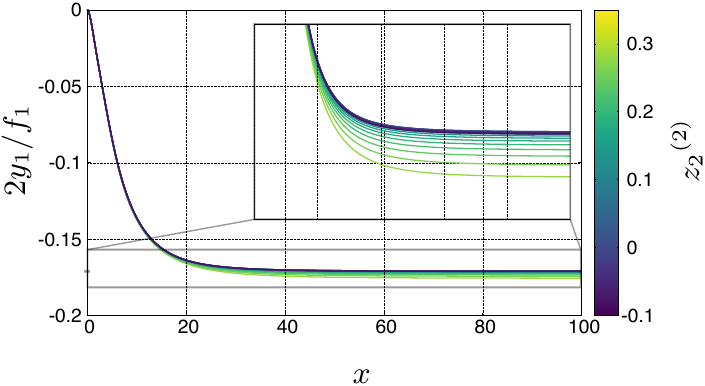}
    \subcaption{The function $2y_1/f_1$}
    \label{fig:tuningDragging}
  \end{subfigure}
  \caption{
    ${z_2}^{(2)}$ dependence of the functions $f_1/x^2$ and $2y_1/f_11$ with fixed $\Omega=0.1$, $\beta=0.001$ and $\lambda=10$.
    We can construct solutions with $f_1/x^2 = 0$ at infinity keeping $2y_1/f_1$ finite by tuning ${z_2}^{(2)}$.  
    }
  \label{fig:tuning} 
\end{figure}

Let us check the functions $f_1/x^2$ and $2y_1/f_1$ which characterize the squashing of $S^3$ and the frame-dragging effect, respectively. 
\figref{fig:functions} shows the functions $f_1/x^2$ and $2y_1/f_1$ for ${z_2}^{(2)}=0$, $\beta=0.001$ and $\lambda=10$.
Unlike the $\Omega=0$ case, $S^3$ is squashed at infinity even if we set ${z_2}^{(2)} = 0$. 
This is because of the contribution of the rotating Vlasov matter.
\figref{fig:tuning} shows the ${z_2}^{(2)}$ dependence of the functions $f_1/x^2$ and $2y_1/f_1$ for $\Omega=0.1$, $\beta=0.001$ and $\lambda=10$.
As we can see from \figref{fig:tuningSquash}, we can choose the value of $f_1/x^2$ at infinity by tuning ${z_2}^{(2)}$. 
On the other hand, the value of $2y_1/f_1$ at infinity is kept finite in general although it can be eliminated at infinity by taking the co-rotating frame. 
As in this example, we can always construct an asymptotically AdS solution that is not squashed at infinity with a finite angular momentum by tuning ${z_2}^{(2)}$.

\section{Conclusion}
\label{sec:conclusion}
We have constructed rotating Einstein-Vlasov systems with the $R_t \times SU(2)_{\bm{\xi}} \times U(1)_{\bm{\sigma}}$ isometry group imposing the exponential form of the distribution function as $\sim \exp\left[\beta(\varepsilon-\Omega j_\sigma)\right]$ with 
$\varepsilon$ and $j_\sigma$ being the conserved energy and angular momentum for the particle motion.  
Under this assumption, the system reduces to the thermal equilibrium state in the limit $\Omega \to 0$.
Since we assume the cosmological constant is negative, we can construct stationary solutions with finite mass without any artificial wall due to the confined structure of the AdS potential.
The distribution of the angular momentum density also decays towards infinity because the angular momentum is carried by the Vlasov matter. 
Therefore we can construct the system with finite mass and total angular momentum by putting $\Omega\neq0$ and solving the Einstein equations.

Even if we set $\Omega = 0$, the system has an additional degree of freedom describing the squashing of $S^3$ at infinity. 
We can control the squashing parameter $s$ at infinity to vanish by tuning the free parameter ${z_2^{(2)}}$ for the boundary conditions at the center.
For $\Omega = 0$ cases, we can construct non-squashed solutions by setting ${z_2^{(2)}} = 0$ because the spherical symmetry is trivially kept with ${z_2^{(2)}} = 0$. 
On the other hand, for $\Omega \neq 0$, each spatial hyper-surface of a given value of radial coordinate $r$ 
is non-trivially squashed with ${z_2^{(2)}} = 0$ due to the contribution of the non-spherical distribution of the Vlasov matter. 
Therefore the parameter ${z_2}^{(2)}$ has to be adjusted for the solution with $s=0$ and the spacetime to be asymptotically AdS with finite total angular momentum. 
We have explicitly demonstrated that this fine-tuning is possible for a specific parameter set. 

In this paper, we have assumed that the distribution function of the particles is dependent on only $(\veps,j_\sigma)$ as a simple model.
The system has, however, one more conserved quantity $J_\xi$ so we can construct solutions with $J_\xi$ dependence keeping the system having the same symmetry.
If we assume an appropriate ansatz, we may construct rotating shell solutions with finite total angular momentum with/without a central black hole.
Although it is hard to analyze such a system in a $4$--dimensional spacetime without spherical symmetry, 
similar analyses would be relatively easier for a $5$--dimensional spacetime with respecting the symmetry of $R_t \times SU(2)_{\bm{\xi}} \times U(1)_{\bm{\sigma}}$. 

\section*{Acknowledgments}
We would like to thank K. Murata for his comments and helpful discussions. 
This work was supported in part by JST SPRING, Grant Number JPMJSP2125 (H. A.), and 
in part by JSPS KAKENHI Grant
Numbers JP19H01895~(CY), JP20H05850~(CY), JP20H05853~(CY).  
The author (H. A.) would like to take this opportunity to thank the 
``Interdisciplinary Frontier Next-Generation Researcher Program of the Tokai Higher Education and Research System."

\appendix
  \renewcommand{\theequation}{A.\arabic{equation}}
\setcounter{equation}{0}

\section{Integration over the momentum space}
\label{sec:int_mom}
In this appendix, we derive the expressions~\eqref{eq:components_emt}.
In \ref{subsec:measure}, to obtain expressions available in massless cases, we do not set the particle mass to unity.
On the other hand, in \ref{subsec:integration}, we set $m=1$ because we strict the distribution function to Eq.~\eqref{eq:mjlike_dist}.

\subsection{Integral measure}
\label{subsec:measure}
First, let us rewrite the integral measure in the momentum space:
\begin{align}
  \dd{V_p} = -\frac{16\fcn{\delta}{p^2+m^2}\fcn{\Theta}{\veps-2ah j_\sigma/G}}{r^2\sqrt{F}\e{\nu}\sin\theta} \dd{p_t}\wedge\dd{p_r}\wedge\dd{p_\theta}\wedge\dd{p_\phi}\wedge\dd{p_\psi},
  \label{eq:momentum_measure}
\end{align}
in terms of the conserved quantities.
Since the definition of the total angular momentum $J_\xi$~\eqref{eq:def_J_xi} can be rewritten as
\begin{align}
  ({J_\xi}^2-{j_\sigma}^2)\sin^2\theta = {\pqty{p_\theta\sin\theta}}^2 + (p_\psi-p_\phi\cos\theta)^2,
\end{align}
the variables $(J_\xi,j_\sigma)$ satisfy ${J_\xi}^2 \ge {j_\sigma}^2$.
Introducing the variable $\chi$ defined as  
\begin{align}
  R\cos\chi = p_\theta\sin\theta \qc R\sin\chi = p_\psi -p_\phi\cos\theta,
  \label{eq:def_r_chi}
\end{align}
with $R^2 \coloneqq ({J_\xi}^2-{j_\sigma}^2)\sin^2\theta$, we obtain
\begin{align}
  \dd{p_\theta}\wedge\dd{p_\phi}\wedge\dd{p_\psi}
  &= R\csc\theta\dd{R}\wedge\dd{j_\sigma}\wedge\dd{\chi} \nt
  &= J_\xi \sin\theta \dd{J_\xi}\wedge\dd{j_\sigma}\wedge\dd{\chi}.
  \label{eq:int_measure_jjchi}
\end{align}
Defining alternative quantities as
\begin{align}
  \tilde{\veps} 
  \coloneqq \veps-\frac{2ah}{G}j_\sigma \qc
  \tilde{J}^2 
  \coloneqq {J_\xi}^2 - \frac{a^2h}{G} {j_\sigma}^2,
  \label{eq:def_tilde}
\end{align}
we can simplify the on-shell condition~\eqref{eq:on-shell_condition} as
\begin{align}
  \frac{G}{F}\tilde{\veps}^2 -\pqty{m^2 +\frac{4\tilde{J}^2}{r^2}} = q^2,
  \label{eq:on-shell_tilde}
\end{align}
and the positive energy condition as $\tilde{\veps} \ge 0$.
Using this condition, we can eliminate the delta function.
Solving Eq.~\eqref{eq:on-shell_tilde} with respect to $q \coloneqq \e{\nu}p^r$, we obtain
\begin{align}
  \fcn{q}{\veps,J_\xi,j_\sigma} = \bqty{\frac{G}{F}\pqty{\veps-\frac{2ah}{G}j_\sigma}^2 -\pqty{m^2 +\frac{4}{r^2}\pqty{{J_\xi}^2-\frac{a^2h}{G}{j_\sigma}^2}}}^{\frac{1}{2}}
  \label{eq:def_q}
\end{align}
for positive $q$.
Using the fact that the integrand is an even function of $q$ due to the ansatz of the distribution function \eqref{eq:dist_ansatz}, we obtain
\begin{align}
  -2 \int \fcn{\delta}{p^2+m^2}\dd{p_t}\wedge\dd{p_r} = \frac{2\e{\nu}\dd{\veps}}{\fcn{q}{\veps,J_\xi,j_\sigma}}. 
  \label{eq:measure_eps}
\end{align}
As a result, the integral measure in the momentum space $\dd{V_p}$ becomes
\begin{align}
  \dd{V_p} = \frac{16\fcn{\Theta}{\tilde{\veps}}J_\xi}{r^2\fcn{q}{\veps,J_\xi,j_\sigma}\sqrt{F}}\dd{\veps}\wedge\dd{J_\xi}\wedge\dd{j_\sigma}\wedge\dd{\chi}.
  \label{eq:measure_epsjjchi}
\end{align}
By the definition of $(\tilde{\veps},\tilde{J})$, it can be rewritten as
\begin{align}
  \dd{V_p} = \frac{16\fcn{\theta}{\tilde{\veps}}\tilde{J}}{r^2\sqrt{F}}\bqty{\frac{G}{F}\tilde{\veps}^2 -\pqty{m^2 +\frac{4\tilde{J}^2}{r^2}}}^{-\frac{1}{2}}\dd{\tilde{\veps}}\wedge\dd{\tilde{J}}\wedge\dd{j_\sigma}\wedge\dd{\chi}.
  \label{eq:measure_epstilde}
\end{align}

We consider the domain of tilded variables based on the on-shell condition \eqref{eq:on-shell_tilde}.
Since $j_\sigma$ satisfies ${j_\sigma}^2 \le {J_\xi}^2 = \tilde{J}^2+\frac{a^2h}{G}{j_\sigma}^2$,
the region of $j_\sigma$ becomes:
\begin{align}
  -\frac{\sqrt{G}}{r} \tilde{J} \le j_\sigma \le \frac{\sqrt{G}}{r} \tilde{J}.
  \label{eq:region_j_sigma}
\end{align}
With a fixed value of $\tilde{\veps}$, 
the angular momentum $\tilde{J}$ is bounded by $J_{\mathrm{max}}$:
\begin{align}
  \tilde{J}^2 = \frac{r^2}{4}\bqty{\frac{G}{F}\tilde{\veps}^2-(m^2+q^2)} \le \frac{r^2}{4}\bqty{\frac{G}{F}\tilde{\veps}^2-m^2} \eqqcolon {J_{\mathrm{max}}}^2,
  \label{eq:region_J_tilde}
\end{align}
then we obtain $0\le\tilde{J}\le J_{\mathrm{max}}$.

The variable $\tilde{\veps}$ takes the minimum value $m\sqrt{F/G}$ when $j_\sigma$ and $\tilde{J}$ vanish.
Since $\tilde{\veps}$ does not have an upper bound, the variable $\tilde{\veps}$ 
takes the range 
\begin{align}
  m\sqrt{\frac{F}{G}} \le \tilde{\veps} < \infty.
  \label{eq:region_eps_tilde}
\end{align}
For the massive particles, we can normalize the variable $\tilde{\veps}$ by the rest mass.
Then we can define the variables as
\begin{align}
  \bar{\veps} \coloneqq \sqrt{\frac{G}{F}}\frac{\tilde{\veps}}{m} \qc 
  \bar{J} \coloneqq \frac{\tilde{J}}{J_{\mathrm{max}}} \qc
  \bar{\jmath} \coloneqq \frac{r}{\sqrt{G}} \frac{j_\sigma}{J_{\mathrm{max}}}.
  \label{eq:def_bar_2}
\end{align}
If we treat a massless particle system, we should replace $\tilde{\veps}/m$ with $\tilde{\veps}$ so $\bar{\veps}$ has the dimension of energy.
The integral regions of these variables are
\begin{align}
  \bar{\veps}\in\clop{1}{\infty} \qc \bar{J}\in\clcl{0}{1}, \qand \bar{\jmath}\in\clcl{-\bar{J}}{\bar{J}},
\end{align}
and the integral measure becomes:
\begin{align}
  \dd{V_p}      
  = \frac{2\fcn{\Theta}{\bar{\veps}}k^2\bar{J}}{\sqrt{1-{\bar{J}}^2}}\dd{\bar{\veps}}\wedge\dd{\bar{J}}\wedge\dd{\bar{\jmath}}\wedge\dd{\chi},
\end{align}
where $k\coloneqq\sqrt{\bar{\veps}^2-1}$ is the local kinetic energy of the particle.

\subsection{Performing the integration}
\label{subsec:integration}
We set the particle mass to unity in this subsection.
In our system, the relevant components of $p_\mu p_\nu$ are as follows:
\begin{subequations}
  \begin{align}
    \pqty{p_t}^2 
    &= \frac{F}{G}\pqty{\bar{\veps}^2 +\frac{2ah}{\sqrt{F}}\bar{\veps}k\bar{\jmath} +\frac{a^2h^2}{F}k^2\bar{\jmath}^2}, \\
    p_t p_\phi 
    &= -\veps j_\sigma = -\frac{\sqrt{F}}{2}\pqty{\bar{\veps}k\bar{\jmath} +\frac{ah}{\sqrt{F}}k^2\bar{\jmath}^2}, \\
    \pqty{p_\phi}^2 
    &= {j_\sigma}^2 = \frac{G}{4}k^2\bar{\jmath}^2, \\
    \pqty{p_r}^2 
    &= \e{2\nu}k^2(1-\bar{J}^2), \\
    \gamma^{\mu\nu}p_\mu p_\nu 
    &= 4J_\xi = r^2 \pqty{k^2\bar{J}^2 +\frac{a^2h}{r^2}k^2\bar{\jmath}^2}, \\
    g^{\mu\nu}p_\mu p_\nu
    &= -1
  \end{align}
  \label{eq:components_momentum}%
\end{subequations}
in the coordinate system defined by~\eqref{eq:def_bar_2}.
Similarly, the distribution function can be rewritten as 
\begin{align}
  \fcn{f}{\veps,j_\sigma} = \exp\bqty{\alpha-\bar{\beta}(\bar{\veps}-\bar{\Omega} k\bar{\jmath})}.
  \label{eq:dist_bar}
\end{align}
The integration of~\eqref{eq:components_momentum} is expressed in the form:
\begin{align}
  2\int_1^\infty\dd{\bar{\veps}}\int_0^1\dd{\bar{J}}\int_{-\bar{J}}^{\bar{J}}\dd{\bar{\jmath}}
  \ \bar{\veps}^i k^j \e{-\bar{\beta}\bar{\veps}} 
  \cdot\bar{J}^{m+1}(1-\bar{J}^2)^{\pm\frac{1}{2}}
  \cdot\bar{\jmath}^n\e{\varsigma\bar{\jmath}}, 
  \label{eq:needed_int}
\end{align}
where $\varsigma \coloneqq \bar{\beta} \bar{\Omega} k = \bar{\beta} \bar{\Omega} \sqrt{\bar{\veps}^2-1}$ with integers $m$ and $n$.
To perform the integration over the momentum space, we divide the integrations into the angular momentum sector and the energy sector.
The angular momentum sector in Eq.~\eqref{eq:needed_int} is expressed as
\begin{align}
  \mathcal{J}^\pm_{m,n}(\varsigma) 
  = 2\int_0^1\dd{\bar{J}}\bar{J}^{m+1}(1-\bar{J}^2)^{\pm\frac{1}{2}}\int_{-\bar{J}}^{\bar{J}}\dd{\bar{\jmath}}\ \bar{\jmath}^n\e{\varsigma \bar{\jmath}}. 
  \label{eq:def_J^pm_mn}
\end{align}
For the even number $m$, since the integrand of Eq.~\eqref{eq:def_J^pm_mn} is an even function, it can be rewritten as
\begin{align}
  \mathcal{J}^\pm_{m,n}(\varsigma) 
  &= \int_{-1}^1\dd{\bar{J}}\bar{J}^{m+1}(1-\bar{J}^2)^{\pm\frac{1}{2}}\mathcal{K}_n(\varsigma,\bar{J}),
\end{align}
where $\mathcal{K}_n(\varsigma,\bar J)$ is defined by:
\begin{align}
  \mathcal{K}_n(\varsigma, \bar J) 
  &= \int_{-\bar J}^{\bar J}\dd{\bar{\jmath}}\ \bar{\jmath}^n\e{\varsigma\bar{\jmath}}.
  \label{eq:def_K_n}
\end{align}
Explicit forms of Eq.~\eqref{eq:def_K_n} for some specific values of $n$ are 
\begin{subequations}
  \begin{align}
    \mathcal{K}_0(\varsigma, \bar J) 
    &= \frac{\e{\varsigma\bar J}-\e{-\varsigma\bar J}}{\varsigma},\\
    \mathcal{K}_1(\varsigma, \bar J) 
    &= -\frac{\e{\varsigma\bar J}-\e{-\varsigma\bar J}}{\varsigma^2} +\frac{\bar J(\e{\varsigma\bar J}+\e{-\varsigma\bar J})}{\varsigma},\\
    \mathcal{K}_2(\varsigma, \bar J) 
    &= \frac{(2+\varsigma^2\bar J^2)(\e{\varsigma\bar J}-\e{-\varsigma\bar J})}{\varsigma^3} -\frac{2\bar J(\e{\varsigma\bar J}+\e{-\varsigma\bar J})}{\varsigma^2},
  \end{align}
\end{subequations}
which appear in the expression of the energy-momentum tensor.
In addition, after integration over $\bar{J}$, we obtain
\begin{subequations}
  \begin{align}
    \fcn{\mathcal{J}^{-}_{0,0}}{\varsigma}
    &= \frac{2\pi \fcn{I_1}{\varsigma}}{\varsigma}, \\
    \fcn{\mathcal{J}^{-}_{0,1}}{\varsigma}
    &= \frac{2\pi \fcn{I_2}{\varsigma}}{\varsigma}, \\
    \fcn{\mathcal{J}^{-}_{0,2}}{\varsigma}
    &= \frac{2\pi \pqty{\fcn{I_2}{\varsigma} +\varsigma\fcn{I_3}{\varsigma}}}{\varsigma^2}, \\
    \fcn{\mathcal{J}^{-}_{2,0}}{\varsigma}
    &= \frac{2\pi \pqty{3\fcn{I_2}{\varsigma} +\varsigma\fcn{I_3}{\varsigma}}}{\varsigma^2}, \\
    \fcn{\mathcal{J}^{+}_{0,0}}{\varsigma}
    &= \frac{2\pi \fcn{I_2}{\varsigma}}{\varsigma^2},
  \end{align}
  \label{eq:result_J^pm_mn}%
\end{subequations}
where $I_\nu(\varsigma)$ is the modified Bessel function of the first kind.

The integration for $\veps$ cannot be analytically performed and we perform it numerically when we solve the Einstein equations. 
For notational simplicity, let us define the function $\mathcal{I}_{i,j,\nu}(\bar{\beta},\bar{\Omega})$ as
\begin{align}
  \fcn{\mathcal{I}_{i,j,\nu}}{\bar{\beta},\bar{\Omega}} = \int_1^\infty \dd{\bar{\veps}}\ \bar{\veps}^i \pqty{\sqrt{\bar{\veps}^2-1}}^j I_\nu(\bar{\beta}\bar{\Omega}\sqrt{\bar{\veps}^2-1}) \exp(-\bar{\beta}\bar{\veps}),
  \label{eq:funcIijnu}
\end{align}
where $i$, $j$ and $\nu$ are non-negative integers, $I_\nu(z)$ is the modified Bessel function of the first kind, and 
\begin{align}
  \bar{\beta}(r) \coloneqq \beta\sqrt{\frac{F}{G}}\qc
  \bar{\Omega}(r) \coloneqq \frac{G}{2\sqrt{F}}\pqty{\Omega -\frac{2ah}{G}}\qc
  \label{eq:bar_parameters}
\end{align}
Then, the resulting forms are given as follows:
\begin{subequations}
  \begin{align}
    T_{tt} 
    &= \frac{4\pi^2\e{\alpha}F}{G}\bqty{\frac{\mathcal{I}_{2,1,1}}{\bar{\beta}\bar{\Omega}} +\frac{2ah}{\sqrt{F}}\frac{\mathcal{I}_{1,2,2}}{\bar{\beta}\bar{\Omega}} +\frac{a^2h^2}{F}\pqty{\frac{\mathcal{I}_{0,2,2}}{{\bar{\beta}}^2{\bar{\Omega}}^2}+\frac{\mathcal{I}_{0,3,3}}{{\bar{\beta}\bar{\Omega}}}}}, \\
    T_{t\phi}
    &= -2\pi^2\e{\alpha}\sqrt{F}\bqty{\frac{\mathcal{I}_{1,2,2}}{{\bar{\beta}\bar{\Omega}}} +\frac{ah}{\sqrt{F}}\pqty{\frac{\mathcal{I}_{0,2,2}}{{\bar{\beta}}^2{\bar{\Omega}}^2}+\frac{\mathcal{I}_{0,3,3}}{{\bar{\beta}\bar{\Omega}}}}}, \\
    T_{\phi\phi} 
    & = \pi^2\e{\alpha}G\pqty{\frac{\mathcal{I}_{0,2,2}}{{\bar{\beta}}^2{\bar{\Omega}}^2}+\frac{\mathcal{I}_{0,3,3}}{{\bar{\beta}\bar{\Omega}}}}, \\
    T_{rr}
    &= 4\pi^2\e{\alpha}\e{2\nu}\frac{\mathcal{I}_{0,2,2}}{{\bar{\beta}}^2{\bar{\Omega}}^2}, \\
    \gamma^{\mu\nu}T_{\mu\nu}
    &=4\pi^2\e{\alpha}r^2 \bqty{\pqty{3+\frac{a^2h}{r^2}}\frac{\mathcal{I}_{0,2,2}}{{\bar{\beta}}^2{\bar{\Omega}}^2} +\pqty{1+\frac{a^2h}{r^2}}\frac{\mathcal{I}_{0,3,3}}{{\bar{\beta}\bar{\Omega}}}}, \\
    g^{\mu\nu}T_{\mu\nu}
    &= -4\pi^2\e{\alpha}\frac{\mathcal{I}_{0,1,1}}{{\bar{\beta}\bar{\Omega}}}.
  \end{align}
  \label{eq:components_emt}%
\end{subequations}
Performing the integral with respect to $\bar{\veps}$ at each $r$, we can obtain the local expression of the energy-momentum tensor.

\renewcommand{\theequation}{B.\arabic{equation}}
\setcounter{equation}{0}

\section{Asymptotic solutions around the center}
\label{sec:bc}

In this Appendix, we investigate the asymptotic behavior around the center in detail and obtain the asymptotic solutions with Vlasov matter.
That is, we determine the coefficients of metric functions in Eq.~\eqref{eq:expand} recursively and confirm that ${z_2}^{(2)}$ is a free parameter in our system.

Substituting Eqs.~\eqref{eq:expand} to Eqs.~\eqref{eq:bar_parameters}, we obtain 
  \begin{subequations}
    \begin{align}
      &\bar{\beta}(x) \simeq 
      \beta_c\bqty{1 +\frac{{u}^{(2)}}{4}x^2 +\frac{{u}^{(3)}}{12}x^3 +\order{x^4}}, \\
      &\bar{\Omega}(x) \simeq 
      \frac{1}{2} \Omega_c x\bqty{1 -\frac{u^{(2)} +w^{(2)}}{4} x^2} +\order{x^4},
    \end{align}%
  \end{subequations}
where $\beta_c \coloneqq \beta\sqrt{{y_3}^{(0)}}$, 
  $\Omega_c \coloneqq \Omega/\sqrt{{y_3}^{(0)}}$, 
  \begin{align}
    {u}^{(i)} \coloneqq \frac{{y_3}^{(i)}}{{y_3}^{(0)}} \qand
    w^{(i)} \coloneqq \frac{4{z_1}^{(i)}}{\Omega_c \sqrt{{y_3}^{(0)}}} -{z_2}^{(i)}
  \end{align}
are constants.
Then we have  
\begin{align}
  I_\nu(uk) \exp(-\bar{\beta}\bar{\veps})
  \simeq \frac{\e{-\beta_c \bar{\veps}}}{\Gamma(1+\nu)}\pqty{\frac{\beta_c \Omega_c}{4} kx}^\nu\bqty{1 +\frac{\zeta_2}{4}x^2 +\frac{\zeta_3}{12}x^3 +\order{x^4}}
\end{align}
with
\begin{align}
  \zeta_2 \coloneqq \frac{{\beta_c}^2{\Omega_c}^2}{4(1+\nu)}k^2 -\beta_c u^{(2)} \bar{\veps} -\nu w^{(2)} \qc
  \zeta_3 \coloneqq -\beta_c u^{(3)} \bar{\veps} -\nu w^{(3)}.
\end{align}
Therefore we can obtain the asymptotic expression as
\begin{align}
  \mathcal{I}_{i,j,\nu}
  \simeq \frac{1}{\Gamma(1+\nu)}\bqty{\frac{\beta_c\Omega_c}{4}x}^\nu \int_1^\infty \dd{\bar{\veps}}\ \bar{\veps}^i k^{j+\nu} \e{-\beta_c \bar{\veps}}\bqty{1 +\frac{\zeta_2}{4}x^2 +\frac{\zeta_3}{12}x^3 +\order{x^4}}.
\end{align}
Using this asymptotic expression, the central value of $T_{tt}$ becomes 
\begin{align}
  T_{tt}^c = \frac{4\pi^2\e{\alpha}(2+\beta_c)(6 +3\beta_c +{\beta_c}^2)}{{\beta_c}^5}{y_3}^{(0)}\e{-\beta_c}. 
\end{align}
Thus we can expand the matter sector in the Einstein equations as
\begin{align}
  s_{\mu\nu} = \sum_{n=0}^\infty \frac{s_{\mu\nu}^{(n)}}{n!}x^n \qc 
  \gamma^{\mu\nu}s_{\mu\nu} = \sum_{n=0}^\infty \frac{s_{\gamma}^{(n)}}{n!}x^n \qc
  g^{\mu\nu} s_{\mu\nu} = \sum_{n=0}^\infty \frac{s^{(n)}}{n!}x^n.
\end{align}
We can calculate the relevant non-vanishing components of them as
\begin{subequations}
  \begin{align}
    s_{tt}^{(0)}
    &= \bqty{\frac{2\pqty{18 + \beta_c \pqty{18 + \beta_c \pqty{7 + \beta_c} } } }{3\pqty{2 + \beta_c}\pqty{6 + \beta_c\pqty{3 + \beta_c}}{y_3}^{(0)}} +\frac{4}{\lambda^2}} {y_3}^{(0)}, \\
    s_{rr}^{(0)}
    &= \frac{9 + \beta_c \pqty{9 + \beta_c \pqty{4 + \beta_c}} }{3\pqty{2+\beta_c}\pqty{6 + \beta_c\pqty{3 + \beta_c}}{y_3}^{(0)}} -\frac{4}{\lambda^2}, \\
    s^{(0)}
    &= \frac{2 {\beta_c}^2 \pqty{1 + {\beta_c} }}{3\pqty{2+\beta_c}\pqty{6 + \beta_c\pqty{3 + \beta_c}}{y_3}^{(0)}} -\frac{20}{\lambda^2}, \\
    s_{t\phi}^{(2)}
    &= -\frac{15+\beta_c(15+6\beta_c+{\beta_c}^2)}{2(2+\beta_c)(6 +3\beta_c+{\beta_c}^2)\sqrt{{y_3}^{(0)}}}\Omega_c,
  \end{align}%
\end{subequations}
and $s_{\gamma}^{(2)} = 12s_{\phi\phi}^{(2)} = 3s_{rr}^{(0)}$ at the leading order. 
Substituting these expressions into the Einstein equations and solving them order by order, we obtain
\begin{align}
  {z_1}^{(2)} &= -\frac{15+\beta_c(15+6\beta_c+{\beta_c}^2)}{6(2+\beta_c)(6 +3\beta_c+{\beta_c}^2)\sqrt{{y_3}^{(0)}}}\Omega_c , \\
  {y_3}^{(2)} &= \frac{18 + \beta_c \pqty{18 + \beta_c \pqty{7 + \beta_c} } }{3\pqty{2 + \beta_c}\pqty{6 + \beta_c\pqty{3 + \beta_c}}} +\frac{2}{\lambda^2}{y_3}^{(0)}, \\
  {y_4}^{(2)} &= {z_2}^{(2)} + \frac{1}{3{y_3}^{(0)}} -\frac{2}{\lambda^2},
\end{align}
at the leading order.
The value of ${z_2}^{(2)}$ is not fixed in this expression and is regarded as a free parameter. 
All other coefficients can be determined by the equations recursively.
Therefore, the boundary conditions of the system are Eq.~\eqref{eq:bc_coefficients} and the parameters are $({y_3}^{(0)},{z_2}^{(2)},\beta,\Omega,\lambda)$ as a whole.

\bibliographystyle{unsrt}
  \bibliography{citation.bib}

\end{document}